\documentstyle[12pt]{report}
\begin{document}

\baselineskip 14pt

\setcounter{chapter}{1}
\setcounter{equation}{0}

\newcommand{\be}{\begin{equation}}
\newcommand{\ee}{\end{equation}}
\newcommand{\n}{\noindent}
\newcommand{\ba}{\begin{eqnarray}}
\newcommand{\ea}{\end{eqnarray}}
\newcommand{\f}{\frac}

\vspace{0.5cm}

\begin{center}
{\Large{\bf On the Aharonov-Bohm Hamiltonian}}
\vspace{2cm}

{\bf R. Adami}\\

S.I.S.S.A., Trieste, Italy.\\

\vspace{0.5cm}

{\bf A. Teta}\\

 Dipartimento di Matematica,\\
Universita' di Roma "La Sapienza", Italy.

\end{center}

\vspace{3cm}

\begin{center}

{\bf Abstract} 

\end{center}

{\small Using the theory of self-adjoint extensions, we construct all the 
possible hamiltonians describing the non relativistic Aharonov-Bohm 
effect. In general the resulting hamiltonians are 
not rotationally invariant so that the angular momentum is not a constant of 
motion. Using an explicit formula for the resolvent, we describe the 
spectrum and compute the generalized eigenfunctions and the scattering 
amplitude.}

\newpage

{\bf 1. Introduction}

\vspace{0.5cm}

\n
In this paper we discuss  the dynamics  of a non relativistic, spinless quantum particle  interacting with a magnetic 
field confined in a thin, infinite solenoid.

\n
Ignoring the irrelevant coordinate along the solenoid, the problem reduces 
to two dimensions and,  if the radius of the solenoid goes to zero while the 
flux of the magnetic field is kept constant, one has  
a particle moving in $R^{2}$ 
subject to a $\delta$-like magnetic field.

\n
At a classical level such a  particle has a trivial dynamics while the quantum 
mechanical description reveals a non trivial scattering cross section, 
explicitely depending on the flux of the magnetic field.

\n
Then one has a purely quantum effect (the Aharonov-Bohm effect ([AB])) due to 
the non local character of the wave function, which is still  
debated in the literature both in its experimental and theoretical 
aspects (see e.g. [R] and references therein).

\n
Here we shall concentrate on the mathematical problem of the construction 
of the hamiltonian describing the quantum dynamics.

\n
Taking the origin of the coordinate system at the position of the solenoid
 and introducing the vector potential 
$A(x,y)= - \alpha ce^{-1} (- y(x^2 + y^2)^{-1}, x(x^2 + y^2)^{-1})$, 
where $-2 \pi c \alpha e^{-1}$ is the magnetic flux through the solenoid, 
the formal hamiltonian of the particle written in polar coordinates reads

\be
\hat{H}_{\alpha} = - \f{\partial^2}{\partial r^2} - \f{1}{r} 
\f{\partial}{\partial r} + \f{1}{r^2} \left( i \f{\partial}{\partial \phi}
- \alpha \right)^{2}.
\ee

\n
In (1.1) we have fixed $\hbar =1$, $m=1/2$; moreover, without loss of 
generality, we suppose $0<\alpha<1$.

\n
A natural starting point for the  construction of  a self-adjoint (s.a.) hamiltonian 
 is to define (1.1) on a domain of smooth functions vanishing for $r=0$, 
e.g. $C_{0}^{\infty}(R^2 \setminus \{0\})$, and then  look for the possible 
s.a. extensions.

\n
Each of the extensions will be characterized by a specific behaviour, i.e. 
boundary condition, of the elements of the domain near $r=0$.

\n
If one requires regularity in $r=0$   for the elements of the domain of  
the extension, one obtains the hamiltonian studied in [AB].

\n
Using the partial wave expansion, the hamiltonian can be reduced to each 
subspace of fixed angular momentum and a complete set of eigenfunctions can 
be constructed, i.e. the model is explicitely solvable.

\n
In recent years it has been realized ([GHKL],[GMS],[MT]) that, if one drops the 
assumption of regularity in $r=0$, a new two-parameter  group of s.a. 
extensions can be constructed.

\n
Such extensions are obtained analysing (1.1) reduced separarely to the subspaces of 
angular momentum zero (s-wave) and minus one (p-wave).These extensions 
obviously commute  with  angular momentum.

\n
Explicit computation of the corresponding new bound states and scattering 
cross sections have also been given.

\n 
These hamiltonians should be interpreted as describing the magnetic 
interaction plus a contact interaction of the particle with the solenoid, 
in analogy with the contact or point interactions perturbing the free 
laplacian ([AGH-KH]).

\n
The relevant difference is that in the latter case  a point interaction can 
only be defined in the s-wave, while in our case the presence of the 
solenoid makes it possible to construct a point interaction also in the 
p-wave.

\n
In this paper we generalize the previous results, developping further the 
work contained in [A].

\n
We give a complete 
description of all the possible s.a. extensions of (1.1) defined on 
$C_{0}^{\infty}(R^{2} \setminus \{ 0 \})$.

\n
We shall make use of the standard theory of s.a. extensions of Von Neumann 
and Krein (see e.g. [RSII],[AG]).

\n
Since  the deficiency indices are $(2,2)$,  there is a 
 family of s.a. extensions of (1.1), parametrized by the unitary map $U$ 
from one deficiency subspace to the other. Since the subspaces have 
 two dimensions, the parametrization involves four real parameters.

\n
The interesting point is that in general  $U$ realizes a coupling 
between the s- and p-waves and then $H^{U}_{\alpha}$ is not  
rotationally invariant, i.e. the angular momentum is not a constant of  
motion.

\n
Only for special values of the parameters in $U$ the previous, rotationally 
invariant extensions, are obtained.

\n
To our knowledge, this new vorticity effect produced by the solenoid 
has not been discussed in the literature.

\n
We shall also discuss the spectral properties of $H^{U}_{\alpha}$. In 
particular we shall analize the occurrence of bound states and we shall 
prove asymptotic completeness. Using the eigenfunction expansion we shall  give rather explicit 
formulas for the scattering amplitude.

\n
A final comment  concerns an alternative description of the 
Aharonov-Bohm problem and its relation with the problem of anyons, i.e.  
particles in $R^{2}$ satisfying fractional statistics (see e.g. [W] and 
references therein).

\n
It is well known that a unitary map reduces (1.1) to the free laplacian 
with boundary conditions on the positive x-axis. The existence of the 
dynamics for the pure magnetic interaction is then proved using the methods 
of potential theory. The construction   generalizes to the case of $N$ solenoids
([S],[DFT]), but  in this approach it seems more difficult to introduce 
the richer structure produced by the contact interactions discussed here.

\n
The hamiltonian  (1.1) is also unitarily equivalent to the hamiltonian 
for the relative motion of two anyons. Our results can then be read as  the 
construction of the most general hamiltonian for two anyons with point 
interaction, the new feature being that  
 such point interaction is not necessarily 
rotationally invariant.

\n
The general N-anyons problem with two-body point interactions seems 
to be difficult to control  with the approach proposed here. A more natural 
setting is probably the one discussed in [DFT].

\n
The remaining part of the paper is organized as follows.

\n
In Section 2 we give the construction of  $H^{U}_{\alpha}$.

\n
In Section 3 we derive an explicit formula for the resolvent and describe 
the spectrum.

\n
In Section 4 we compute the generalized eigenfunctions and the scattering 
amplitude.

\vspace{1cm}

\setcounter{chapter}{2}
\setcounter{equation}{0}

{\bf 2. Self-adjoint extensions}

\vspace{0.5cm}

\n
As stated in Sec. 1, we start considering the symmetric and positive 
operator $\hat{H}_{\alpha}$ given by (1.1) with $D(\hat{H}_{\alpha})=
C^{\infty}_{0}(R^{2} \setminus \{0\} )$, in the Hilbert space 
$L^{2}(R^{2})$. We denote its closure by $\dot{H}_{\alpha}$, where

\be
D(\dot{H}_{\alpha}) = \left\{ u \in L^{2}(R^{2}) \;|\; u \in 
H^{2}_{loc}(R^{2} \setminus \{ 0 \}), \; \hat{H}_{\alpha} u \in
L^{2}(R^{2}) \right\}.
\ee

\n
$H^{n}$ (resp. $H^{n}_{loc}$) is the standard Sobolev space (resp. local 
Sobolev space) of order $n$.

\n
To construct all the s.a. extensions of $\dot{H}_{\alpha}$ we consider 
 the solution of the equation 
$\dot{H}_{\alpha}^{*} \psi = \pm i  \psi$, where $\dot{H}_{\alpha}^{*}$ 
is the adjoint of $\dot{H}_{\alpha}$.

\n
This  equation is more conveniently studied if we introduce the 
decomposition of the Hilbert space $L^{2}(R^{2})$ with respect to angular 
momentum

\be
L^{2}(R^{2}) = L^{2}(R^{+}; rdr) \otimes L^{2}(S^{1}),
\ee

\n
where $R^{+} = (0 ,\infty)$ and $S^{1}$ is the unit sphere in $R^{2}$. 
Using the unitary transformation

\be
V \; : \; L^{2}(R^{+};rdr) \rightarrow L^{2}(R^{+}) \;\;\;\; 
(Vf)(r) = r^{1/2} f(r)
\ee

\n
and the completeness  of $\frac{e^{im \phi}}{\sqrt{2 \pi}}$, $m \in 
Z$, in $L^{2}(S^{1})$, we can also write

\be
L^{2}(R^{2}) = \bigoplus_{m = - \infty}^{+ \infty}  
\left( V^{-1} L^{2}(R^{+}) \right) \otimes 
\left[ \frac{e^{im \phi}}{\sqrt{2 \pi}} \right],
\ee

\n
where $[ \frac{e^{im \phi}}{\sqrt{2 \pi}}]$ is the linear span of
$\frac{e^{im \phi}}{\sqrt{2 \pi}}$. 
Corresponding to the decomposition (2.4) one has

\be
\dot{H}_{\alpha} = \bigoplus_{m=- \infty}^{+ \infty}
V^{-1} \dot{h}_{\alpha,m} V \otimes 1
\ee

\n
where the operators $\dot{h}_{\alpha,m}$ in $L^{2}(R^{+})$ are defined 
by (see e.g. [AGH-KH])

\ba
& &D(\dot{h}_{\alpha,0}) = \left\{ \xi \in L^{2}(R^{+}) \;|\; \xi,\xi' 
\in H^{1}_{loc}(R^{+}), \right. \nonumber \\
& &\left. -\xi'' + (\alpha^{2} - 1/4) r^{-2} \xi 
\in L^{2}(R^{+}), \; W(\xi, \xi^{(0)}_{\pm})_{0^{+}}=0 \right\},
\ea

\ba
& &D(\dot{h}_{\alpha,-1}) = \left\{ \xi \in L^{2}(R^{+}) \;|\; \xi,\xi'
\in H^{1}_{loc}(R^{+}), \right. \nonumber\\
& &\left. -\xi'' + [(1- \alpha)^{2} - 1/4] r^{-2} \xi
\in L^{2}(R^{+}), \;  W(\xi, \xi^{(-1)}_{\pm})_{0^{+}}=0 \right\}
\ea

\ba
& &D(\dot{h}_{\alpha,m}) = \left\{ \xi \in L^{2}(R^{+}) \;|\; \xi,\xi'
\in H^{1}_{loc}(R^{+}), \right. \nonumber\\
& &\left. -\xi'' +  [(m+ \alpha)^{2} - 1/4] r^{-2} 
\xi \in L^{2}(R^{+}) \right\}, \;\;\;\; m \neq 0,-1,
\ea

\be
\dot{h}_{\alpha,m}\xi =- \frac{d^{2}\xi}{dr^{2}}  + \frac{(m + \alpha)^{2}
- 1/4}{r^{2}}\xi, \;\;\;\; m \in Z
\ee

\n
Here $W(f,g)_{x} = \overline{f(x)} g'(x) - \overline{f'(x)} g(x)$ denotes the 
Wronskian of $f$ and $g$ evaluated in $x$ and 

\be
\xi^{(0)}_{+}(r) = N r^{1/2} K_{\alpha} \left( e^{-i \frac{\pi}{4}} r 
\right), \;\;\;\; \xi^{(0)}_{-}(r) = N e^{i \frac{\pi}{2}\alpha}  
r^{1/2} K_{\alpha} \left( e^{i \frac{\pi}{4}} r \right),
\ee

\be
\xi^{(-1)}_{+} (r) = M r^{1/2} K_{1-\alpha} \left( e^{-i \frac{\pi}{4}}
r \right), \;\;\;\; \xi^{(-1)}_{-}(r)= M e^{i \frac{\pi}{2} (1- \alpha)}
r^{1/2} K_{1-\alpha} \left( e^{i \frac{\pi}{4}} r \right),
\ee

\be
N= \frac{\sqrt{2 \cos \frac{\pi}{2} \alpha}}{\pi}\; , \;\;\;\; 
M= \frac{\sqrt{2 \sin \frac{\pi}{2} \alpha}}{\pi}.
\ee

\n
$N$ and $M$ are normalization factors and $K_{\nu}$ is 
 the McDonald function of order $\nu$ 
([GR]). The phase factors in $\xi^{(0)}_{-}$, $\xi^{(-1)}_{-}$ are  
irrelevant at this step. They have been introduced  only to be consistent 
with the choice of an analytic basis for the deficiency subspaces of 
$\dot{H}^{*}_{\alpha}$ (see below).

\n
It is well known (see e.g. [RSII],[BG]) that $\dot{h}_{\alpha,m}$ are s.a. for 
$m \neq 0, -1$, while $\dot{h}_{\alpha,0}$ and $\dot{h}_{\alpha,-1}$ have 
deficiency indices (1,1).

\n
The adjoint operators of $\dot{h}_{\alpha,0}$ and $\dot{h}_{\alpha,-1}$ 
are defined by

\ba
& &D(\dot{h}^{*}_{\alpha,0}) = 
\left\{ \xi \in L^{2}(R^{+}) \;|\; \xi, \xi' \in H^{1}_{loc}(R^{+}),\right.
\nonumber\\
& &\left. - \xi'' + (\alpha^{2} - 1/4) r^{-2} \xi \in L^{2}(R^{+}) \right\},
\ea

\ba
& &D(\dot{h}^{*}_{\alpha,-1}) =
\left\{ \xi \in L^{2}(R^{+}) \;|\; \xi, \xi' \in H^{1}_{loc}(R^{+}),\right.
\nonumber\\
& &\left.- \xi'' + [(1 - \alpha)^{2} - 1/4] r^{-2} \xi \in L^{2}(R^{+}) \right\},
\ea

\be
\dot{h}^{*}_{\alpha,m} \xi = -\frac{d^{2} \xi}{dr^{2}} + 
\frac{(m + \alpha)^{2} - 1/4}{r^{2}}
\xi, \;\;\;\; m=0,-1.
\ee

\n
Moreover (2.10) and (2.11) are the unique solutions (up to a constant factor) of 
$\dot{h}^{*}_{\alpha,0} \zeta = \pm i \zeta$ and $\dot{h}^{*}_{\alpha,-1} 
\zeta = \pm i \zeta$ respectively.

\n
From (2.4),(2.15) we obtain

\be
\dot{H}^{*}_{\alpha} = \bigoplus_{m=- \infty}^{+ \infty} V^{-1} 
\dot{h}^{*}_{\alpha,m} V \otimes 1\;.
\ee

\n
Using (2.16) we easily find that the equation $\dot{H}^{*}_{\alpha} \psi = 
\pm i \psi$ has two independent solutions

\be
\psi^{(0)}_{\pm} (r) =\frac{r^{-1/2}}{\sqrt{2 \pi}} \xi^{(0)}_{\pm}(r), \;\;\;\;
\psi^{(-1)}_{\pm}(r, \phi) =\frac{r^{-1/2}}{\sqrt{2 \pi}}  \xi^{(-1)}_{\pm}(r) 
e^{-i \phi}.
\ee

\n
This means that $\dot{H}_{\alpha}$ has deficiency indices (2,2) and then a 
four-parameter family of s.a. extensions.

\n
We take $ \left\{ \psi^{(0)}_{\pm}, \psi^{(-1)}_{\pm} \right\}$ as a basis  for the two 
dimensional deficiency subspaces $\chi_{\pm} \equiv ker(\dot{H}_{\alpha}^{*}
 \mp i)$ and denote by $U$ any unitary map from $\chi_{+}$ to $\chi_{-}$. 
Applying the standard  theory ([AG]), one 
 finds that such extensions are explicitely given by

\ba
& &D(H^{U}_{\alpha})= \left\{ u \in L^{2}(R^{2}) \;|\; u = v + \psi_{+} 
+ U \psi_{+} \right\}, \\
& &H^{U}_{\alpha}u = \dot{H}_{\alpha}v + i \psi_{+} - i U \psi_{+},
\ea

\n
where $v \in D(\dot{H}_{\alpha})$, $\psi_{+}= c_{0} \psi^{(0)}_{+} + 
c_{-1} \psi^{(-1)}_{+}$, $c_{0},c_{-1} \in C$, $U\psi_{+} = 
\tilde{c}_{0} \psi^{(0)}_{-} + \tilde{c}_{-1} \psi^{(-1)}_{-}$, 
$\tilde{c}_{j} = \sum_{l=0,-1} \tilde{U}_{jl} c_{l}, \; j=0,-1$, and 
$\tilde{U}$ is a $2 \times 2$ unitary matrix which can be represented as

\be
\tilde{U} = e^{i \eta} \left( \begin{array}{cc}
                               a & -\bar{b} \\
                               b & \bar{a}
                              \end{array} \right), 
\;\;\;\; |a|^{2} + |b|^{2} =1.
\ee

\n
For any possible choice of the parameters $\eta,\; a, \; b$ we obtain a 
s.a.  realization of the formal hamiltonian (1.1). 

\n
If we choose  
$\eta=0$, $b=0$, $a=-1$ the original extension $H^{AB}_{\alpha}$ studied by Aharonov-Bohm is 
obtained. In this case the domain of the hamiltonian consists of functions 
vanishing for $r \rightarrow 0$. The only parameter appearing in the 
hamiltonian is $\alpha$, which correspond to a purely magnetic interaction 
(i.e. 
with no additional contact interaction in $r=0$).

\n
For $b=0$ one has $a=e^{i \tau}$ and then the extensions are parametrized 
by $\tau, \eta$. Such extensions have been studied recently in 
[GHKL],[GMS],[MT]. They correspond to a magnetic interaction plus a contact interaction acting 
in the s-wave and in the p-wave  separately.

\n
The choice $\eta=\pi + \tau$ (resp. $\eta= \pi - \tau$) eliminates  
 the contact interaction acting in the p-wave (resp. in the s-wave).

\n
For $b \neq 0$ one has extensions which don't commute with the 
angular momentum.

\vspace{1cm}

\setcounter{chapter}{3}
\setcounter{equation}{0}
{\bf 3. The Resolvent}

\vspace{.5cm}
\noindent
In this Section we  use Krein's method ([AG],[DG]) to compute the resolvent of the generic self-adjoint extension $H^U_\alpha$.

\noindent
The two basic ingredients are the knowledge of the resolvent of a particular extension of $\dot{H}_\alpha$ and the construction of an analytic basis  for the deficiency subspaces.

\noindent
In Sec. 2 we observed that for $\eta  =  0$, $b  =  0$, and  $a  =   -1$, i.e. $U  =  -I$, one has $H_\alpha^U  =  H_\alpha^{AB}$, where the Aharonov-Bohm extension $H_\alpha^{AB}$ is an exactly solvable hamiltonian.

\noindent
Its spectrum is $[0 \, , \, + \infty )$, it is purely absolutely continuous, and the generalized eigenfunctions are
\begin{equation}
\Psi_\alpha^{AB} \, (r,\phi,k ,\theta) \ = \ \sum_{m =  - \infty}^{+ \infty} \, i^{|m|} \,  {e^{i m (\phi - \theta )}}
\, e^{i \, \frac \pi 2 \, ( |m|  - |m + \alpha |)} \, J_{|m + \alpha |} (kr) \label{autofunzab}
\end{equation}
(see e.g. [R]). $J_\nu$ is the Bessel function of the first kind and of order $\nu$. 

\noindent
From (\ref{autofunzab}) one easily obtains the explicit expression for the resolvent of $H_\alpha^{AB}$. 
\begin{eqnarray}
\left[ {\cal R}_\alpha^{AB} \, (k) \, f \right] \, (r, \phi) & \equiv & \left[ \left(  H_\alpha^{AB}  -  k^2  \right)^{-1} \, f \right]
  (r, \phi) \nonumber \\ & & \nonumber \\
                                 & = & \int_0^{+\infty} \, \rho \, d \rho \, 
\int_{0}^{2 \pi} \, d \zeta \sum_{m \, = \, - \infty}^{+\infty} i 
\, \frac {e^{i m (\phi - \zeta)}} 4 \nonumber \\ & & \times \, 
J_{|m + \alpha|} ( k \, (r \wedge \rho)) \  H_{|m + \alpha|}^{(1)} 
(k \, (r \vee \rho)) \ f (\rho, \zeta) \label{risab}
\end{eqnarray}
where $Im \, k > 0$, $(x \wedge y)$ and $(x \vee y)$ are respectively the minimum and the maximum between $x$ and $y$.
$H_\nu^{(1)}$ is the Hankel function of the first kind and of order $\nu$.

\vspace{.3cm}
\noindent
The construction of an analytic basis can be made using (\ref{risab}).

\noindent
We observe that the following map, defined for $Im \, k > 0$
\begin{eqnarray}
{\cal V} (k,e^{i \frac \pi 4}) & \equiv & (H_\alpha^{AB}  -  i) \, (H_\alpha^{AB}  -  k^2)^{-1} \nonumber \\
   & = & 1 \, +  \, (k^2  -  i) \, {\cal R}_\alpha^{AB} (k)
\end{eqnarray}
is obviously analytic in $k$ and defines an isomorphism between $\chi_+$ and $\chi (k) \equiv Ker (\dot H_\alpha^* - k^2)$.

\noindent
Then the analytic basis in  $\chi (k)$ is:
\begin{equation}
\left\{ \psi_k^{(0)},~ \psi_k^{(-1)} \right\} \ = \ \left\{ {\cal V} (k, e^{i \frac \pi 4})  \,  \psi_+^{(0)}, ~ {\cal V} (k, e^{i \frac \pi 4})  \,  \psi_+^{(-1)} \right\}
\label{quattro} \end{equation} 

\noindent
For $k^2  =  -i$ in (\ref{quattro}) we reobtain the basis $\left\{ \psi_-^{(0)}, \psi_-^{(-1)} \right\}$ for $\chi_-$.

\noindent
Now the generalized Krein's formula ([AG], [DG]) yields

\begin{eqnarray}
{\cal R}_\alpha^U (k) \, f & \equiv & (H_\alpha^U \, - \, k^2)^{-1} \, f   \nonumber \\
                           & = & {\cal R}_\alpha^{AB} (k) \, f \, +  \sum_{j,l = 0, -1}
\ p(k)_{jl} \ (\psi_{- \overline k}^{(j)}\, , \, f) \ \psi_k^{(l)} \label{risolvente}
\end{eqnarray}
Here the $2 \times 2$ matrix $p(k)$ is determined by
\begin{eqnarray}
p(k) & = &
 \left( 1 \, + \, (k^2 - i ) \, p(e^{i \frac \pi 4}) \, A (k, e^{i \, \frac \pi 4}) \right)^{-1} \, p(e^{i \, \frac \pi 4})
\label {pidikappa} \end{eqnarray}
where $A (k_1, k_2)$ is a $2 \times 2$ complex matrix whose elements can be 
written as follows
\begin{eqnarray}
A (k_1, k_2)_{jl} & \equiv &  (\psi_{- \overline k_1}^{(j)}\, , \, \psi_{k_2}^{(l)}) \nonumber
\\ & & \nonumber \\
& = & \frac 1 {k_2^2 \, - \, k_1^2} \left[ \frac {(-k_1^2)^\alpha - (-k_2^2)^\alpha} {sin \, (\frac \pi 2 \alpha)} \ \delta_{j0} \, \delta_{jl} \nonumber \right. \\ 
 & & \left.
+ \  \frac {(-k_1^2)^{1-\alpha} - (-k_2^2)^{1-\alpha}} {cos \, (\frac \pi 2 \alpha)} \ \delta_{jl} \, \delta_{-1l} \right]~~~~~j,l \, = \, 0, -1 \end{eqnarray}
 $p(e^{i \frac \pi 4})$ is a $2 \times 2$ matrix which can be computated
following ([DG]) 
\begin{eqnarray}
p(e^{i \frac \pi 4}) & = & \frac i 2 \ A(e^{i \frac \pi 4}, e^{i \frac 3 4 \pi})^{-1} \ \left( -I \, - \, \overline {\tilde U} \right) \nonumber \\
& & \nonumber \\                     & = & - \frac i 2 \left( 
\begin{array}{cc}
1 \, + \, e^{-i \eta} \, {\overline a}  &  - \, e^{-i \eta} \ b \\ & \\  e^{-i \eta} \ {\overline b} & 1 \, + \, e^{-i \eta} 
 \, a \end{array} \right) \label{otto}
\end{eqnarray} 
We have  denoted in (\ref{otto}) by $\overline B$ the matrix obtained from the matrix $B$ taking the complex conjugate of each element.

\vspace{.3cm}
\noindent

It is easy to show that under the condition $Im \, k > 0$ 
the inverse matrix in (\ref {pidikappa}) exists, except for (at most) two 
values of $k$, corresponding to the bound states of $H_\alpha^U$
 ([DG]).

\vspace{.3cm}
\noindent
A straightforward computation shows that
\begin{eqnarray}
p_{00} (k) & = & \frac {e^{-i \eta}} {2 D} \left[ \frac {a' \, + \, cos \, \eta}{cos \left( \frac \pi 2
 \alpha \right)} \left( (-k^2)^{1 - \alpha} \, - \, (-i)^{1 - \alpha} \right) \, - \, i \, (e^{i \eta} + \overline a )
\right] \nonumber \\ & & \nonumber \\
p_{0 -1}(k) & = & \frac {i \, e^{-i \eta}} {2D} \ b \nonumber \\ & & \nonumber \\
p_{-1 0} (k) & = & - \frac {i \, e^{-i \eta}} {2D} \ \overline b \nonumber \\ & & \nonumber \\
p_{-1-1} (k) & = & \frac {e^{-i \eta}} {2 D} \left[ \frac {a' \, + \, cos \, \eta}{sin \left( \frac \pi 2
 \alpha \right)} \left( (-k^2)^{\alpha} \, - \, (-i)^{\alpha} \right) \, - \, i \, (e^{i \eta} +  a )
\right]
\end{eqnarray}
where
\begin{eqnarray}
D & = & det \left( 1 \, + \, (k^2 - i ) \, p(e^{i \frac \pi 4}) \, A (k, e^{i \, \frac \pi 4}) \right) \nonumber \\
  & = & c_1 (-k^2) \, + \, c_\alpha (-k^2)^\alpha \, + \,  c_{1-\alpha} (-k^2)^{1-\alpha} \, + \, c_0 
\end{eqnarray}
with
\begin{eqnarray}
c_1 & = & - \frac {e^{-i \eta}} {sin \, (\pi \alpha )} \ (a' \, + \, cos \, \eta) \nonumber \\
c_\alpha & = & \frac {e^{-i \eta}} {sin \, (\pi \alpha )} \left[ a' \, sin \, \left( \frac \pi 2 \alpha \right) \,
 + \,  sin \, \left( \frac \pi 2 \alpha \, - \, \eta \right) \, + \, a'' \, cos  \left( \frac \pi 2 \alpha \right)
\right] \nonumber \\
c_{1- \alpha} & = &   \frac {e^{-i \eta}} {sin \, (\pi \alpha )} \left[ a' \, cos \, \left( \frac \pi 2 \alpha \right) \,
 + \, cos  \, \left( \frac \pi 2 \alpha \, + \, \eta \right) \, - \, a'' \,  sin  \left( \frac \pi 2 \alpha \right)
\right] \nonumber \\ 
c_0 & = &  \frac {e^{-i \eta}} {sin \, (\pi \alpha )} \left( sin \, \eta \, - \, a'' \, cos (\pi \alpha)
\, - \, a' \, sin (\pi \alpha) \right) 
\end{eqnarray}
and
\begin{eqnarray}
a' & = & Re \, a \nonumber \\
a'' & = & Im \, a \end{eqnarray}

\noindent
Due to the structure of the resolvent ${\cal R}_\alpha^U (k)$ as given by (\ref{risolvente}), one can easily obtain information on the spectral properties of $H_\alpha^U$ (cfr. the analogous situation for ordinary point interactions in 
[AGH-KH]).

\vspace{.3cm}
\noindent
In particular one has
\begin{equation}
\sigma_{ac} (H_\alpha^U) \ = \ \sigma_{ac} (H_\alpha^{AB}) \ = \ [0, + \infty)~~~~~~\sigma_{sing} (H^U_\alpha) 
\ = \emptyset
\end{equation}

\noindent
Moreover the pure point spectrum consists of (at most)  two negative eigenvalues determined as the solutions of
\begin{equation}
det \left( 1 \, + \, (k^2 - i ) \, p(e^{i\frac \pi 4}) \, A (k, e^{i \, \frac \pi 4}) \right) ~ = ~ 0 \label{statilegati}
\end{equation}

\noindent
Concerning the scattering, one knows (see e.g. [R]) that the wave operators associated to the pair $(H_\alpha^{AB}, -\Delta)$ exist and are complete in the strongest sense
\begin{equation}
Ran \, \left( \Omega_\alpha^{AB} \right)_+ \ = 
\  Ran \, \left( \Omega_\alpha^{AB} \right)_- \ = 
\ {\cal H}_{bound \; states}^\perp
\end{equation}
and  the eigenstates have negative energy.

\noindent
Using the chain rule for the wave operators and the Kato-Birman theorem
([RS III, SIM]), we conclude that also the wave operators for the pair $(H_\alpha^U, -\Delta)$ exist and are complete in the strongest sense.

\vspace{.3cm} \noindent
We observe that  eq. (\ref{statilegati}) is easily analyzed in the special case $\eta = 0$, $a = 0$, $b = e^{i \gamma}$,
corresponding to the simplest, non-rotationally invariant hamiltonian.

\noindent
In fact for $k^2 = - |E|$, eq. (\ref{statilegati}) now reads
\begin{equation}
- \, |E| \, + \, sin \left( \frac \pi 2 \alpha \right) \, |E|^\alpha \, + \,  cos \left( \frac \pi 2 \alpha \right) \, |E|^{1-\alpha} \ = \ 0 \label{lesempio}
\end{equation}
It is easy to verify that (\ref{lesempio}) has two solutions: 
$|E| = 0$, corresponding to a zero-energy resonance, and $|E| 
= E_1 (\alpha)$ corresponding to a  non rotationally invariant bound state.

\noindent
Finally, one can also note that for the  rotationally-invariant case 
$( b = 0 , ~ a = e^{i \tau})$ equation (3.14) becomes 
\begin{equation}
\left[ (-k^2)^{1 -\alpha} \,  cos \, \omega \,  - \, sin \, \left(
\frac \pi 2 \alpha \, - \, \omega \right) \right] \,
\left[ (-k^2)^\alpha \,  cos \, \beta \, - \, cos \, \left( \beta 
+ \frac \pi 2 \alpha \right) \right]  =  0 
\end{equation}
where $\omega \, = \, \frac {\eta - \tau} 2$ and $\beta \, = \, \frac
{\eta + \tau} 2$.

\noindent
From this factorization one  obtains the equations for the bound states in the 
p- and  s-wave respectively (cfr. [A], [MT]).

\vspace{1cm}

\setcounter{chapter}{4}
\setcounter{equation}{0}

{\bf 4. Stationary Scattering Theory}

\vspace{.5cm}
\noindent
Here we use the above results to describe the stationary scattering theory for the pair $(H_\alpha^U, -\Delta)$.

\noindent
In particular we compute the generalized eigenfunctions and the scattering amplitude for $H_\alpha^U$.

\noindent
A complete (time-dependent) analysis of the scattering for the couple 
$(H_\alpha^U, -\Delta)$  requires further 
technical work and  will be discussed in a forthcoming paper.

\noindent
Denote by $R_\alpha^U (k; \rho, \theta, r, \phi)$, the integral kernel of the resolvent of $H_\alpha^U$.

\vspace{.3cm}
\noindent
Then the generalized eigenfunctions $\Psi_\alpha^U(k, \theta, r, \phi)$, with 
$k \geq 0$, of $H_\alpha^U$ are obtained through the standard limit procedure
(see e.g. [AG], [AGH-KH])
\begin{eqnarray}
\Psi_\alpha^U (k, \theta, r, \phi) & = & \lim_{\varepsilon \downarrow 0} \lim_{\rho \rightarrow + \infty} \,
\frac 4 {i \, H_0^{(1)} ((k + i \varepsilon ) \rho )} \ R_\alpha^U (k+i 
\varepsilon ; \rho, \theta + \pi, r, \phi) \\
 & & \nonumber \\ & & \nonumber \\
  & = &   \Psi_{\alpha}^{AB}(k, \theta,r,\phi) \nonumber \\
  & & + \, 2 \, i^{\alpha + 1} \, cos \left( \frac \pi 2 \alpha \right) \, p_{00} (k) \, (-k^2)^\alpha \, H_\alpha^{(1)} (kr) 
\nonumber \\ & & \nonumber \\
  & & - \, i^{\alpha } \, \sqrt{2 \, sin (\pi \alpha)} \, e^{-i \pi (1 -2 \alpha)} \, p_{-10}(k) \, k \, H_\alpha^{(1)} (kr) \,
e^{i \theta} \nonumber \\ & & \nonumber \\
  & & + \, i^{1 - \alpha} \,  \sqrt{2 \, sin (\pi \alpha)} \, e^{i \pi (1 -2 \alpha)} \, p_{0-1}(k) \, k \, H_{1-\alpha}^{(1)} (kr) \,
e^{-i \phi} \nonumber \\ & & \nonumber \\
  & & - \, 2 \, i^{2 - \alpha} \, sin \left(\frac \pi 2 \alpha \right) \, p_{-1-1}(k) \, (-k^2)^{1-\alpha} \, H_{1-\alpha}^{(1)}
(kr) \, e^{-i (\phi - \theta)} \nonumber \\
\end{eqnarray}

\vspace{.3cm} \noindent
Finally we determine the scattering amplitude $f_\alpha^U (k, \theta, \phi)$ from the asymptotic behaviour of $\Psi_\alpha^U
(k, \theta, r, \phi)$ for large $r$, i.e.
\begin{equation}
\Psi_\alpha^U (k, \theta, r, \phi) \ \stackrel {r \rightarrow +\infty} \longrightarrow \ e^{ikr \, cos (\phi - \theta)} \, + \,
f_\alpha^U (k, \theta, \phi) \, \frac {e^{ikr}} {\sqrt r}
\end{equation} 
where
\begin{eqnarray}
f_\alpha^U (k, \theta, \phi) & = & f_\alpha^{AB} (k, \theta, \phi) \nonumber \\ & & \nonumber \\
  & & + \, 4 \, \sqrt{\frac {2i\pi} k} \, cos \left( \frac \pi 2 \alpha \right) \, p_{00} (k) \, (-k^2)^\alpha \nonumber \\
& & \nonumber \\
  & & -  \, 2 \, \sqrt{\frac {2\pi} {ik}} \,  \sqrt{2 \, sin (\pi \alpha)} \, e^{-i \pi (1 -2 \alpha)} \, p_{-10}(k) \, k \,
e^{i \theta} \nonumber \\ & & \nonumber \\ 
  & & +   \, 2 \, \sqrt{\frac {2\pi} {ik}} \,   \sqrt{2 \, sin (\pi \alpha)} \, e^{i \pi (1 -2 \alpha)} \, p_{0-1}(k) \, k \, e^{-i \phi} \nonumber \\ & & \nonumber \\
  & & - \, 4 \, \sqrt{\frac {2i\pi} k} \, sin \left( \frac \pi 2 \alpha \right) \, p_{-1-1} (k) \, (-k^2)^{1-\alpha} e^{-i (\phi - \theta)} \label{ampiezza}
\end{eqnarray}

\vspace{.3cm} \noindent
We have denoted by
$f_\alpha^{AB}$  the scattering amplitude associated to $H_\alpha^{AB}$
explicitely given by (see e.g. [R]):
\begin{eqnarray}
f_\alpha^{AB} & = & \sqrt \frac {2 \pi} {ik} \left[ \delta ( \phi - \theta)\, (cos (\pi \alpha) - 1) \,  + \, i \frac {sin (\pi \alpha)}
\pi \, P \frac 1 {e^{-i (\phi - \theta)} -1} \right]
\end{eqnarray}

\vspace{.3cm} \noindent
The above formula for  the scattering amplitude seems 
to be of interest. For $b=0$ it shows that 
the scattering process can be reduced
 in each subspace  of fixed angular  momentum.

\noindent
On the other hand, for $b \neq 0$, i.e., 
for any extension which is not 
rotationally invariant, one 
obtains from  (\ref{ampiezza}) a non 
vanishing and computable probability 
 for the scattering process:  incoming particle with angular momentum zero 
(resp. minus one) $\longrightarrow$ outcoming particle
with angular  momentum minus one (resp. zero); the probability is 
proportional
to $|p_{0-1}(k)|^2$ (resp. $|p_{-10}(k)|^2$).

\noindent
The lack of conservation of  angular momentum for such 
hamiltonians can therefore be directly checked on a 
physical observable quantity,  the scattering 
cross section for a given scattering process.

\newpage

{\bf Acknowledgements}

\vspace{.4cm}
\noindent
We thank prof. G.F.
Dell'Antonio for constant encouragement and helpful discussions
 during the entire preparation of this paper.

\vspace{.4cm} \noindent
During the final draft of the paper we became aware that L. Dabrowski and P. Stovicek 
were concluding a preprint on the same subject ([DS]).

\vspace{1cm}

{\bf References}

\vspace{0.5cm}

\noindent
[A] R. Adami, {\it  Tesi di Laurea, Universita' di Pisa}, July 1996 (in 
Italian).

\vspace{.3cm}
\noindent
[AB]        Y. Aharonov and D. Bohm, Phys. Rev. {\bf 115} (1959) 485.

\vspace{.3cm}
\noindent
[AG]    N. Akhiezer and I. Glazman, {\em Theory of Linear Operators in Hilbert Space},  Frederich Ungar, New 
York, (1963).

\vspace{.3cm}
\noindent
[AGH-KH] S. Albeverio, F. Gesztesy, R. H\o gh-Krohn and H. Holden, 
{\em Solvable Models in Quantum Mechanics}, Springer-Verlag, New York 
(1988).

\vspace{.3cm}
\noindent
[BG] W. Bulla, F. Gesztesy, J. Math. Phys. {\bf 26} (1985) 2520-2528.

\vspace{.3cm}
\noindent
[DFT] G.F. Dell'Antonio, R. Figari, A. Teta, to appear in Lett. Math. Phys. (1997).

\vspace{.3cm}
\noindent
[DG] L. Dabrowski, H. Grosse, J. Math. Phys. {\bf 26} (1985) 2777-2780.

\vspace{.3cm}
\noindent
[DS] L. Dabrowski, P. Stovicek, preprint SISSA - Trieste, December 1996.

\vspace{.3cm}
\noindent
[GHKL] J. Grundberg, T.H. Hausson, A. Karlhede, J.M. Leinaas, 
Mod. Phys. Lett. B, {\bf 5-7} (1991) 539-546.

\vspace{.3cm}
\noindent
[GMS]  P. Giacconi, F. Maltoni and R. Soldati, preprint  DFUB/95 - 6 
(1995).

\vspace{.3cm}
\noindent
[GR]  I.S. Gradshteyn and I.M. Ryzhik, 
{\em Table of Integral Series and Products}, Academic Press, San Diego 
(1979).

\vspace{.3cm}
\noindent
[MT] C. Manuel and R. Tarrach, Phys. Lett. B {\bf 268} (1991) 222-226.

\vspace{.3cm}
\noindent
[R]  S.M.N. Ruijsenaars, Ann. of Phys., {\bf 146} (1983) 1-34.

\vspace{.3cm}
\noindent
[RS II]  M. Reed and B. Simon, {\em Methods of Modern Mathematical 
Physics}, vol.2, {\em Fourier 
Analysis and Self-Adjointness}, Academic Press (1978).

\vspace{.3cm}
\noindent
[RS III]     M. Reed and B. Simon, {\em Methods of Modern 
Mathematical Physics}, vol.3, {\em Scattering Theory}, Academic Press 
(1979).

\vspace{.3cm}
\noindent
[S] P. Stovicek, Proc. of the Workshop on Singular Schr\"odinger Operators, 
Trieste 1994, preprint  ILAS/FM  16/1995.

\vspace{.3cm}
\noindent
[SIM]   B. Simon, {\em Quantum Mechanics for Hamiltonian Defined as Quadratic Forms}, Princeton University Press, 
New York (1971).

\vspace{.3cm}
\noindent
[W] F. Wilczek, Phys. Rev. Lett. {\bf 49} (1982) 957-1149.

\end{document}